# Identifying User Behavior in domain-specific Repositories

Wilko VAN HOEK[a,1], Wei SHEN[a] and Philipp MAYR[a]

[a] *GESIS – Leibniz Institute for the Social Sciences, Germany*

**Abstract.** This paper presents an analysis of the user behavior of two different domain-specific repositories. The web analytic tool *etracker* was used to gain a first overall insight into the user behavior of these repositories. Moreover, we extended our work to describe an apache web log analysis approach which focuses on the identification of the user behavior. Therefore the user traffic within our systems is visualized using chord diagrams. We could find that recommendations are used frequently and users do rarely combine searching with faceting or filtering.

## Introduction

In recent years, many web analytics applications have been published to measure and analyze usage data in order to understand and optimize the information seeking in web systems. When designing a domain-specific repository, it is important to understand the ways in which users perform searches. A lot of studies were conducted to understand user behavior in the context of web search analysis. Bates proposed a dynamic search model, describing that searcher's information needs change over time [2]. She further extended her work by characterizing the common information seeking process that consists of sequences of search tactics [1]. To investigate the human information search process, Koch et al. [3] conducted a thorough log analysis, which grouped the session-based log entries into eleven different activities and used these activities to identify user behavior. Mayr [5] presented a quantitative, non-reactive measure for standard Apache log files focusing on typical navigation types which can easily be extracted from the referrer information in the log.

Domain-specific repositories always target at a certain user group, which has substantial domain expertise and aims to search for specialized, domain-oriented information. Typically specialized users develop individual search tactics [1] to operate in the repositories in an efficient way. The traffic of these users leave traces in the web server log which can be consulted for detailed analyses of navigational structures of a system. Russell-Rose et al. [6] categorized users into four types: double experts, domain expert/technical novices, domain novice/technical experts and double novices. In this sense, we assume that users of domain-specific repositories could be ranked as double experts or domain expert/technical novices. Double experts are individuals identified with high domain and technical expertise, which often use teleporting search strategy, formulating the queries precisely and jump quickly to the destination [7]. Domain expert/technical novices, on the other hand, are able to use their knowledge to formulate effective queries, but lack the technical confidence to explore unknown territory [4].

---

[1] Corresponding author: Wilko van Hoek, GESIS – Leibniz Institute for the Social Sciences, Unter Sachsenhausen 6-8, 50667 Cologne, Germany; E-mail: wilko.vanhoek@gesis.org

In this paper, we report preliminary findings of a user behavior analysis within three domain-specific collections. The three collections belong to two different repositories. Two of the three collections are part of the *Effektiv!*[2] portal and the third collections is the *Social Science Open Access Repositories (SSOAR)*[3].

## 1. Background

The *Effektiv!* portal is an academic online portal funded by the German Federal Ministry of Education and Research which offers descriptions of programs to support family friendliness at German education institutions and disclosed best practices to help the scholars and students to balance better between an academic career and their family life. The core parts of the *Effektiv!* portal are two collections. An online database with practice examples of family-friendly best practices in academic education institutions (herein after called *Effektiv! best practices*) and a bibliography of literature specialized in family-friendliness and gender topics (herein after called *Effektiv! literature*). Both collections are online since April 2013.

Founded in 2008 the *SSOAR* is a full-text server for open access publications in the field of social sciences. Furthermore, *SSOAR* offers the social scientists, scientific associations and publishers the opportunity to self-archive their publications, to enhance the visibility of their work on the web. There are currently about 27,600 digital papers archived in *SSOAR*.

The portals *SSOAR* and *Effektiv!* are both based on the same repository software DSpace. Different search user interfaces have been designed to help the user to apply specific search strategies. A guided search concept is applied to design the *Effektiv! literature* and *Effektiv! best practices* user search interface. The main design idea behind the *Effektiv! literature* search interface is: besides an overall search box, the user can enter further search terms for assumed popular attributes such as author and title in additional search boxes. Users are further invited to select values of two additional filters. A browsing of the "subject area" is provided at the right side of the site. In the *Effektiv! best practices* user search interface, filters are emphasized and presented at the top of the search form while the standard search box is at the bottom. In this case, users are encouraged to narrow down their searches quickly with the selection of these filters. On the contrary, the main search interface for *SSOAR*, called *browse and search*, is designed with a faceted search concept, in which attributes are displayed as links in a navigational menu. This approach facilitates the user to intuitively search by progressively refining their choices. In addition to the faceting, users can browse the system by disciplines. When a discipline is selected, the user can apply facets or search in the result list. In this way facetted search and browsing can be combined. Besides the *browse and search* interface, a traditional advanced search is also provided, supporting users to freely formulate their search queries.

---

[2] Effektiv! - For Greater Family Friendliness in German Higher Education Institutions (www.familienfreundliche-hochschule.org). Funded by the German Federal Ministry of Education and Research (BMBF) (grant no. 01FW11101). Any opinions expressed here are those of the author(s).

[3] http://www.ssoar.info

In the following chapters we want to gain insights into the way the different search options are used in the three collections. We want to find out which concepts work well and which do not.

## 2. Web Analysis Using *etracker*

The *etracker*[4] web analysis software was used to identify the general user behavior of the two repositories. The investigated time period is from 1st April 2013 and 31st December 2013.

According to *etracker* there were 254,240 users visiting the SSOAR portal and 5,641 users visiting the *Effektiv!* portal during this time period. We examined the user number, page impression, visiting time of the user interface in each repository as shown in Table 1. We can see that the most SSOAR users (over 90% of all) went to the *browse and search* user interface to search for documents. The advanced search interface was rarely used. About 14% of the *Effektiv!* portal visitors used the best practices collection and only 5% of visitors viewed the literature collection. This effect may be due to the structure and multi-functionality of the *Effektiv!* portal. Besides the two collections, the *Effektiv!* portal also provides services like online advisory service, press information etc., which means that the main goal of many *Effektiv!* portal visitors may not be the best practices or literature search.

The parameters "page impressions per user" and the visiting time were calculated in both SSOAR search interfaces. However, the *Effektiv!* best practices user interface was apparently much more viewed than the literature user interface. And although the visiting time per page of best practices was adjacent, the users within the best practices user interface took more time visiting than the users within the literature. This may indicate that compared to *Effektiv! literature* collections, the *Effektiv!* users made more search queries in the *Effektiv! best practices* collection.

**Table 1:** Summary statistics of different user interfaces

| Repository name/ User Interface name | Total users | Page impressions | Page impressions per user | Visiting time per user | Visiting time per page |
|---|---|---|---|---|---|
| *Effektiv! best practices* | 788 | 4,100 | 5.20 | 00:02:27 | 00:00:28 |
| *Effektiv! literature* | 331 | 1,219 | 3.68 | 00:01:55 | 00:00:31 |
| *SSOAR browse and search* | 24,421 | 117,765 | 4.82 | 00:03:29 | 00:00:43 |
| *SSOAR advanced-search* | 3,616 | 15,574 | 4.31 | 00:03:00 | 00:00:42 |

The click path chart indicates the user's movement paths through the web pages. Figure 2 shows the click path chart of the *SSOAR browse and search* user interface (here *SSOAR/discover/*). The yellow node in the middle of the chart represents the discovery user interface. The grey *entry* node above represents the direct entry in the user interface from other pages (compare approach in [5]), while the grey *exit* node beneath represents requests where users left the page. The five top ranked referrer sites are listed at the left side of the graph and the five top following sites are listed at the right side. About 25% of the users went from *SSOAR* homepage (German and English) to the *browse and search* interface and 16.8% came directly from external pages. Nearly 30% of the users left the *SSOAR* portal after viewing the interface. Due to the fact that over 50% of the sites are not analyzed and grouped to the *others* node, this click path analysis in *etracker* is very limited and the sequential search process cannot be thoroughly displayed.

---

[4] www.etracker.com

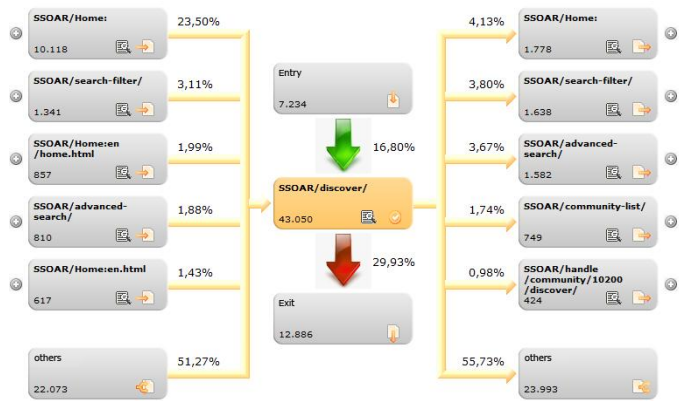

**Figure 1:** Click path chart

## 3. Apache Web Log Analysis

Although the analysis of the *etracker* data has given us some first insights of the users' behaviors, many details are missing. We don't know which types of interaction are most or least frequently used. The main problem seems to be that *etracker* cannot identify what type of action is performed on a page and what kind of information lies behind an URL. For instance, in the click tracker analysis most of the traffic has been grouped together as *others*. It is not possible to figure out in *etracker* how many document views, searches or browsing actions the group *others* are consisting of.

To overcome this lack of detail, we decided to analyze the raw log files of our web servers. On both systems we are using the Apache 2[5] web server with an identical logging configuration. During the time period between 1st April 2013 and 31st December 2013 we collected IP (anonymized), timestamp, requested URL and referrer of all visitors were collected. As the functionality of both systems relies mainly on http-requests, we can identify what page is viewed by analyzing the URLs given in the log.

To understand the users' behavior, we focused on analyzing the pairwise information of requested URL and referrer. So to say we looked at where users were coming from and where they were going to. Both systems are using the software Solr[6] as their search backend. This allows us to identify what pages where requested by analyzing the URL. For instance, we can see if a simple search, based on a single query, is conducted or whether a more complex search, using filters or facets has been executed. We grouped the user traffic into different types of search interactions (see table 2 and table 3) and then calculated how many requests involved users to switch between these types.

---

[5] Apache Server Project (http://httpd.apache.org/)

[6] Apache Solr (http://lucene.apache.org/solr/)

When analyzing web server log files, it is important to clean out automated accesses e.g. by web spiders that usually generate the biggest amount of traffic. Spiders systematically request every part of a web page. Most of those spiders can be identified by their IP address. The software DSpace collects lists of spiders. We used these lists to clean out all know web spiders. In addition we truncated all requests regarding the hour at which the access was conducted and counted the number of request per IP address. Based on this data we could identify a small set of further IP addresses responsible for a large amount of traffic. We then filtered out the data generated by those IP addresses.

In the following subsections, we will describe our analysis of the web server log files. We will introduce a new visualization technique applied for log data. Thereafter we will present the results of our analysis for both *Effektiv!* collections and *SSOAR*.

## 3.1. Chord diagrams

In the following sections, we use chord diagrams[7] to visualize the traffic within the three collections (*SSOAR*, *Effektiv! literature* and *Effektiv! best practices*). We used the D3.js library[8] to create the diagrams. These diagrams have originally been used to visualize the movements of people between different neighborhoods[9]. We transferred this idea to our situation by interpreting different types of search interactions as neighborhoods and the transition from one to another as a movement between two neighborhoods. For example, using free *text search* is one type of interaction and assigning filters another. When a user changes the result list for a free text search by applying filters, we counted this as a transition between two neighborhoods. At first we defined two types of web pages that can be used in all interfaces (cf. table 2). Then we identified the different types of search interactions (cf. table 3 and table 4).

**Table 2:** General page types accessed by users

| Page type | Effektiv! literature & best practices | SSOAR |
|:---:|:---:|:---:|
| G1 | Initial search page | Initial search page |
| G2 | Document URL | Document URL |

**Table 3:** Search interaction types performed by users

| Search interaction type | Effektiv! literature & best practices | SSOAR |
|:---:|:---:|:---:|
| S1 | List-all | Repository overview |
| S2 | Free text search without filter | Free text search without facet |
| S3 | Filter search | Faceted search |
| S4 | Free text search with filter | Free text search with facet |
| S5 | Change query | Change facet |
| S6 | - | Advanced-search |

---

**Table 4:** Browsing interaction types performed by users

| Browsing interaction type | Effektiv! literature & best practices | SSOAR |
|---|---|---|
| B1 | Browsing | Browsing |
| B2 | - | Browsing with free text search |
| B3 | - | Browsing with facet |
| B4 | - | Browsing with free text search and with facet |

To understand the way chord diagrams work we will give an example. Figure 2 illustrates the traffic related to the *Effektiv! literature* collection. The chord diagram can be read as follows. The total amount of data is represented by a circle. The data is grouped around this circle. Each type of interaction or page is represented by an arc. The size of an arc represents the amount requests where the referrer URL was assigned to that type of interactions or pages. The area between two arcs illustrates the traffic between the corresponding types. For instance, in approx. one third of the requests where the referrer was assigned to browsing (type B1), the destination URL belongs to a document (type G2). In reverse, most of the traffic where the referrer is a document URL (type G2) has a destination URL that was assigned to browsing (type B1).

### 3.2. Log file analysis results for *Effektiv! literature*

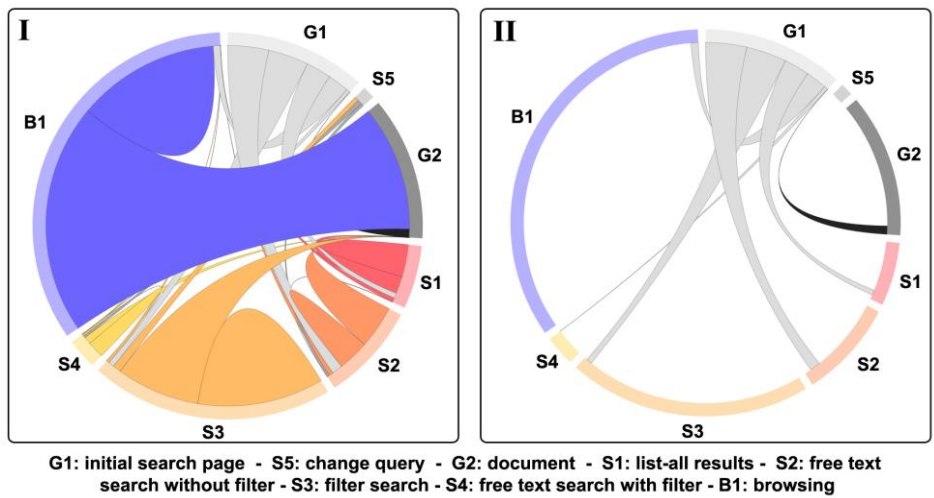

G1: initial search page  -  S5: change query  -  G2: document  -  S1: list-all results  -  S2: free text search without filter  -  S3: filter search  -  S4: free text search with filter  -  B1: browsing

**Figure 2:** User traffic for different interaction types for the *Effektiv! literature* database – overview in I and II shows the traffic only for the initial search page (S1).

For the collection *Effektiv! literature* the search interaction browsing (type B1) is responsible for the highest amount of user traffic, as 35% of the referrers hold URLs belonging to browsing. This is followed by filter search (type S3) that takes up 21% of the traffic. Part II of Figure 2 shows which type of interactions users have used after the initial search page (type G1). Overall the users seem to continue quite equally with the different type from the initial search page. Simple search without search terms and searching without filtering are most often used in this situation. Combining search terms with filters is an exception as it is rarely the next step taken by the majority of users. After selecting a way of querying the users only rarely change their way of searching, there are only small amounts of requests in which users change the search terms or switch from one type to another. Also only 10% of the traffic consists of requests where a search term has been entered.

Comparing a filter search and browsing, a basic difference in the users' behavior can be observed. Users who are filtering without query terms are viewing a document in 37% of the cases, but remain within the same search type in 55%. Staying in the same search type means that users are viewing a second result page or sorting their results. When browsing, the users access documents in 60% of the cases and stay in browsing in 37%. Users seem to find relevant documents by browsing more often than by just filtering. The second observation in this context is that many requests show movements from documents to browsing. This can be explained by links that are shown on a document page. For instance, users can proceed from document pages by browsing the system using the author name. This may also be an explanation for the dominance of the browsing related traffic in the log files.

### 3.3. Log file analysis results for *Effektiv! best practices*

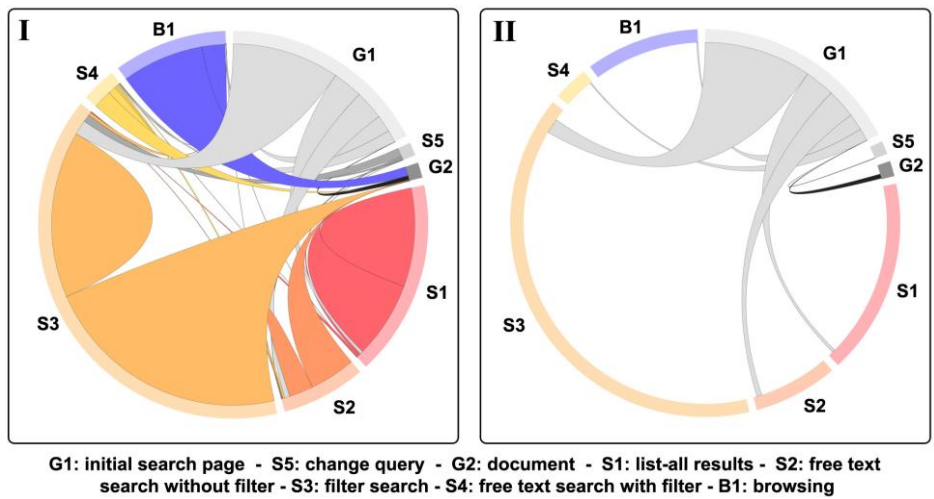

**G1: initial search page  -  S5: change query  -  G2: document  -  S1: list-all results  -  S2: free text search without filter  -  S3: filter search  -  S4: free text search with filter  -  B1: browsing**

**Figure 3:** User traffic for different interaction types for the *Effektiv! best practices* collection – overview in I and II shows the traffic only for node A.

Figure 3 illustrates the user traffic related to the *Effektiv! best practices* collection. Here filtering without entering search terms (type S3) is the most often requested type with 45% of the total traffic. Listing all results (type S1) and browsing (type B1) are ranked second and third with 16% and 9% of the traffic respectively. The preference to filter without search terms can also be observed by looking at the traffic from the initial search page (type G1). 55% of the users are proceeding from the initial search page, by filtering the data without entering search terms. Overall users tend not to change or reformulate their search query as only a small amount of traffic is related to those cases and users rarely query the system using search terms.

When looking at the relation between users moving from browsing to documents or users moving from filtering without search terms and to documents, it can be observed that the users' behavior is slightly different. 48% of the users that are filtering without search terms access documents while 32% stay within the interaction type. In contrary 77% of the traffic that was generated by users browsing the data led to documents, while only 21% of the users remain browsing. Users that are browsing the system seem to access documents more frequently than users that don't, although more users are filtering the system.

*3.4.* Log file analysis results for the SSOAR

In SSOAR there are two interfaces that allow the user to search in the collection. There is the advanced-search that allows querying by searching in specific metadata fields as well as searching over all fields and there is the browse and search interface in which search terms can be combined with facets. The browse and search also allows users to browse for documents and to search the result list of the browsing or apply facets to filter that result list. The browsing functionality has been discussed strongly during the development of the browse and search interface. We therefore decided to distinguish between faceted search (types S1-S5) and browsing (types B1-B4). Figure 4 shows the traffic between the interaction types advanced-search (type S6), faceted search, and browsing.

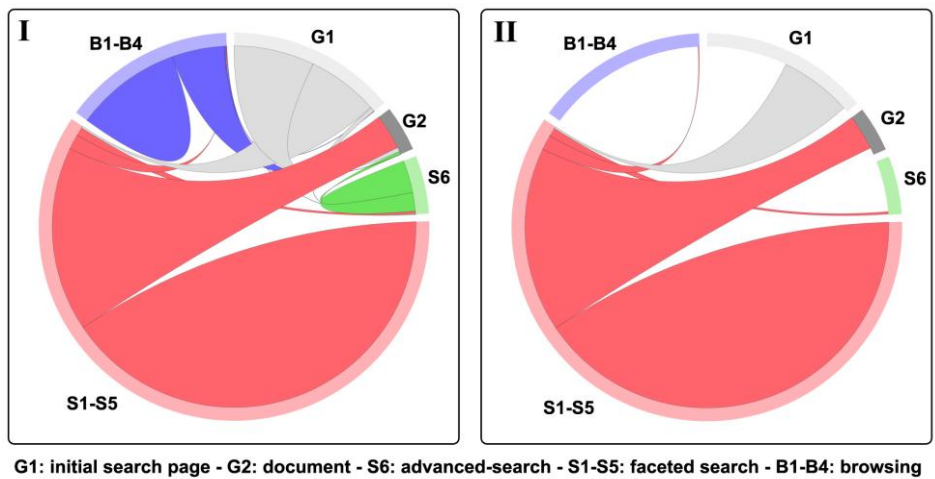

G1: initial search page - G2: document - S6: advanced-search - S1-S5: faceted search - B1-B4: browsing

**Figure 4:** User traffic for different interaction types for SSOAR – overview in I and II shows the traffic only for node D.

62% of the users' traffic is concentrated on faceted search. And 60% of the movement from this interaction type is self-directed. This means that in those cases users conducted interactions like query reformulation or selecting facets. Nearly one third of the group's traffic is related to requests from faceted search to documents (type G2). The next largest amount of traffic for one type, with roughly 20 % of the total traffic, is browsing. Here similar to faceted search, 60% of the movement is self-directed and 40% represents movements to documents. The same observation holds for the advanced-search. When looking at the movement between the three search types, it becomes clear that only a small amount of users switch from faceted search to either browsing or advanced-search and an even smaller proportion of users request the opposite direct. The high amount of traffic from document to faceted search can be explained by the "more about" link. This link is shown on the document page and directly triggers a faceted search using metadata fields such as authors.

To better understand the users' behavior within the faceted search and the browsing; we generated two additional diagrams that show the traffic related to those two types. Interestingly the two do not interact strongly. Users that decide to first browse the system usually remain in that status. This is understandable as users are able to further query and facet in the filtered results and may not be interested in broaden their results again. Figure 5 shows the traffic for faceted search (part I) and browsing (part II).

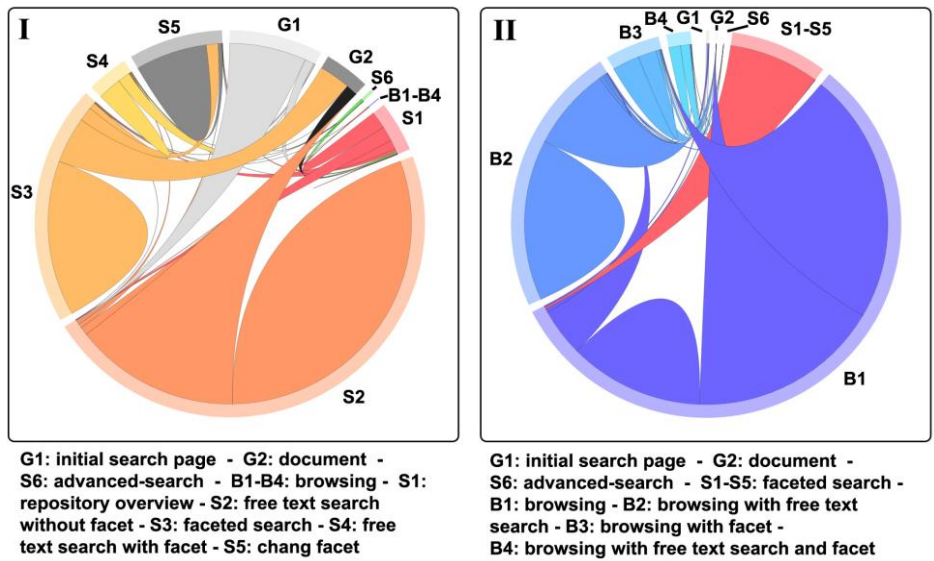

**I**
G1: initial search page - G2: document -
S6: advanced-search - B1-B4: browsing - S1:
repository overview - S2: free text search
without facet - S3: faceted search - S4: free
text search with facet - S5: chang facet

**II**
G1: initial search page - G2: document -
S6: advanced-search - S1-S5: faceted search -
B1: browsing - B2: browsing with free text
search - B3: browsing with facet -
B4: browsing with free text search and facet

**Figure 5:** Different types of user traffic in SSOAR related to faceted search (I) and browsing (II)

Most of the traffic related to faceted search is generated by users that query the system by entering search terms without using facets (50%). In addition this is also the most often used starting point as 88% of the users proceed with this type after the starting page. The second highest amount of traffic belongs to search where only facets are applied (21%). Obviously users prefer to query the system by using search terms or use facets but rarely combine both.

A different situation can be observed when looking at the traffic related to browsing. Browsing without entering search terms not using facets is the most dominant type here with 60% of the traffic. It is followed by browsing with search terms without using facets (24%). An exception of the analyzed data lies in the movement from browsing without search terms to using facets. A high proportion of 40% of the traffic related to browsing without search terms and without facets is related to applying facets. But in total browsing without search terms with facets takes only 5% of the traffic. This way of querying the system seems to be a dead end that we cannot explain right now.

### Conclusion

In this paper, we have presented the results of two analyses of user behavior in the repositories *Effektiv! literature*, *Effektiv! best practices* and SSOAR. In the first analysis we tried to identify user tactics by looking at the information provided by the service *etracker*. In the second analysis we conducted an own evaluation of the raw log files generated by the web servers.

During the first analysis it became clear that the information provided by services like *etracker* do not suffice to identify the user's behavior. We therefore decided to evaluate the traffic generated by our users by ourselves, in form of a log file analysis. Based on our own log files analysis we could observe that users searching in the *Effektiv! literature* collection use the browsing and filtering opportunities intensely and rarely type in search terms. In addition we could see that the links provided on the document pages, where users can proceed within the system by browsing for authors or topics were used frequently.

For the *Effektiv! best practices* collection browsing is less frequently used but filtering is therefore used more often. The difference in the users' behavior in this collection to the *Effektiv! literature* collection can be explained by the fact that the provision of the browsing links presented on the document pages are less interesting to the users. To improve the search in this collection we will consider adding further browsing opportunities.

In the results for SSOAR we could see that advanced-search and browsing is less frequently used than faceted search. Looking into the detailed behavior for faceted search and browsing we could see that users do rarely combine facets and search terms. We consider to change our front end to improve this. The next observation is that faceting when browsing seems to be a dead end decision. We will need to examine this more closely to understand why this is the case. The third major observation is that, similar to *Effektiv! literature*, the opportunity to continue the search from a document by clicking on a link which allows searching for more documents with the same metadata entry is used frequently. We should improve our functionality regarding this feature in SSOAR.

Overall analyzing the log files is worthwhile and should be done more frequently. After setting it up it can be used regularly to analyze the effects frontend changes of a web site result in. Right now we are able to better understand how our systems are used. By improving our method and extending it to identify user-session, we are confident to be able to identify user search tactics and thus gain more information about our users. We will continue to analyze our data regularly in the future.